\documentclass[pra,amsmath,amssymb,showpacs,superscriptaddress,twocolumn]{revtex4}
\input epsf.tex
\usepackage{graphicx}
\usepackage{amsthm}
\usepackage{array}
\usepackage{longtable}
\usepackage{dcolumn}
\usepackage{bm,bbm}
\usepackage[usenames,dvipsnames]{color}
\usepackage{amsmath}
\usepackage[hypertex]{hyperref}
\usepackage{amsfonts}
\usepackage{amssymb}
\usepackage{mathrsfs}
\usepackage{subfigure}
\usepackage{ulem}

\begin{document}    

\title{Effects of systematic phase errors on optimized quantum random-walk search algorithm}

\affiliation {Zhengzhou Information Science and Technology Institute, Zhengzhou, 450004, China}
\affiliation {Synergetic Innovation Center of Quantum Information and Quantum Physics, University of Science and Technology of China, Hefei, Anhui 230026, China}

\author {Zhang Yu-Chao} \author {Bao Wan-Su }\email{2010thzz@sina.com} \author {Wang Xiang}  \author {Fu Xiang-Qun}
\affiliation {Zhengzhou Information Science and Technology Institute, Zhengzhou, 450004, China}
\affiliation {Synergetic Innovation Center of Quantum Information and Quantum Physics, University of Science and Technology of China, Hefei, Anhui 230026, China}

\date{\today}

\begin{abstract}
This paper researches how the systematic errors in phase inversions affect the success rate and the number of iterations in optimized quantum random-walk search algorithm. Through geometric description of this algorithm, the model of the algorithm with phase errors is established and the relationship between the success rate of the algorithm, the database size, the number of iterations and the phase error is depicted. For a given sized database, we give both the maximum success rate of the algorithm and the required number of iterations when the algorithm is in the presence of phase errors. Through analysis and numerical simulations, it shows that optimized quantum random-walk search algorithm is more robust than Grover's algorithm.
\end{abstract}

\pacs{03.67.Dd, 03.67.Hk}
\maketitle

\section{Introduction}  
Quantum computation follows the laws of quantum mechanics such as quantum interference, superposition and entanglement to achieve a new computing pattern of information processing with a powerful parallel computing capability. In 1996, the Grover algorithm$^{[1]}$ proposed by Grover is one of the most typical quantum algorithms. It only needs $O(\sqrt{{{2}^{n}}})$   computational complexity to complete a ${{2}^{n}}$-sized database search, achieving quadratic speed-up compared with the classical exhaustive algorithm. Grover's algorithm makes people see the great power of quantum algorithms. Since then further research based on Grover's algorithm is proposed continuously.$^{[2,3,4]}$ Grover's algorithm has also been proved to be physically implemented.$^{[5]}$

In practice, errors in the implementation of phase inversion are inevitable, including systematic errors and random errors. The error of each step may be very small, but the scale of quantum computation is usually exponential level. So, it may lead to big errors finally and affect the algorithm results. The robustness of the algorithm to the error refers to the stability when it is affected by error interference. It can be measured by the change of success rate before and after the introduction of errors. In 2000, Gui Lu Long found systematic errors in phase inversions and random errors in Hadmard-Walsh transformations are the dominant effects in Grover's algorithm.$^{[6]}$ In 2003, Shenvi studied the robustness of Grover's algorithm to a random phase error in the oracle and the error tolerability of the algorithm.$^{[7]}$ It has a significant value in studying the robustness of quantum algorithms to errors, which can improve the reliability of quantum algorithms and benefiting the implementation of algorithms.

In recent years, quantum random walk has become a research hot topic. Quantum random walk is the counterpart of the classical random walk. It evolves with quantum state and is affected by the quantum interference. In 2003, Shenvi proposed a quantum random-walk search algorithm (SKW algorithm).$^{[8]}$ It iterates $\frac{\pi }{4}\sqrt{{{2}^{n+1}}}$  times and the success rate is 0.5, which is the first algorithm using quantum random walk to reach quadratic speed-up on an unstructured ${{2}^{n}}$-sized database. In 2008, Potocek put forward optimized quantum random-walk search algorithm (optimized SKW algorithm)$^{[9]}$ on the basis of the former. It iterates $\frac{\pi }{4}\sqrt{{{2}^{n}}}$ times and the success rate is 1 approximately, achieving the same  calculation results of Grover's algorithm. Since then, some quantum algorithms based on quantum random walks have been proposed.$^{[10,11,12,13]}$

In 2006, Li Yun studied the effects of errors in phase inversions on SKW algorithm$^{[14]}$ and found that in addition to reducing the probability of marked state eventually, the error also reduces the maximum probability of the unmarked state, which is different from Grover's algorithm. While in the relevant experiments (such as NMR), as long as the probability of the marked state is much higher than other states, the algorithm is still successful. One of the main differences between SKW algorithm and Grover's algorithm is the use of the Grover diffusion operator, $G=2\left| {{S}^{C}} \right\rangle \left\langle  {{S}^{C}} \right|-I$. The Hilbert space of SKW algorithm is the tensor product
${\rm H} = {{\rm H}^C} \otimes {{\rm H}^S}$, where ${{\rm H}^C}$  is the ${n}$ dimensional space associated with the quantum coin, and ${{\rm H}^S}$ is the ${{2}^{n}}$  dimensional Hilbert space representing the vertex. The Grover diffusion operator as the coin operator only acts on the ${n}$ dimensional coin space, thus, $\left| {{S}^{C}} \right\rangle =\frac{1}{\sqrt{n}}\sum\limits_{d=1}^{n}{\left| d \right\rangle }$. But in Grover's algorithm, it acts on the ${{2}^{n}}$  dimensional database, corresponding to the vertex space in SKW algorithm, thus, $\left| {{S}^{C}} \right\rangle =\frac{1}{\sqrt{{{2}^{n}}}}\sum\limits_{x=1}^{{{2}^{n}}}{\left| x \right\rangle }$. The increment of work space will increase the difficulty of quantum gates implementation and design costs. Also, it may result in introducing more errors, causing more interferences to the algorithm. From a practical point of view, the search algorithm based on random walk may have more advantages over Grover's algorithm. Therefore, its in-depth study has important theoretical significance and practical value. Currently, there is no research about the effects of errors on optimized SKW algorithm.

This paper studies the effects of systematic phase errors produced by the Grover diffusion operator in optimized SKW algorithm. With theoretical analysis and numerical simulations methods, we establish the model of the algorithm with phase errors and depict the relationship between the success rate of the algorithm, the database size, the phase error and the number of iterations. For a given sized database, the relationship between the maximum success rate of the algorithm and the phase error is given. The relationship between the required number of iterations and the phase error is also given. A critical value of the database scale is obtained. The algorithm will be almost unaffected by the error if the database scale is smaller than that. Numerical simulations show that optimized SKW algorithm is more robust than Grover's algorithm.

The rest of this paper is organized as follows. In Section 2, we introduce optimized SKW algorithm. The geometric description of optimized SKW algorithm is proposed in Section 3. Section 4 establishes the model of the algorithm with phase errors. In Section 5, we give the results of numerical simulation and make a robustness comparison with Grover's algorithm. The conclusion is summarized in Section 6.
\section{Optimized SKW algorithm}  
\label{sec:1}
Optimized quantum random walk search is based on a quantum random walk on a hypercube. The ${n}$ dimensional hypercube is a graph with ${2^n}$ vertices, each of which can be labelled by an n-bit binary string. Two vertices are connected by an edge if their Hamming distance is 1. The data in the ${2^n}$-sized database is expressed in binary, $\vec x = ({x_1},{x_2} \cdots {x_n})$, ${x_i} = \{ 0,1\} $. $|\vec x|$ is its Hamming weight and $p(\vec x) = |\vec x|\bmod 2$ is its parity bit. Then they can be mapped to a ${2^{n + 1}}$ dimensional space, $\vec x \to \vec x' = \vec x||p(\vec x)$, i.e. adding the parity bit at last. As the oracle contains information of the marked state and it is one-to-one with $\vec x'$, we suppose the oracle is still able to identify the marked state after mapping.

When $\vec x'$ corresponds to the vertex of a dimensional $n + 1$ hypercube, the Hamming weights of these vertices are even. Assuming that the marked vertex is $\left| 0 \right\rangle $, its selection does not affect the final result since the hypercube is symmetrical. This database search problem is then equivalent to searching for a single marked vertex amongst the even vertices on the hypercube.

In detail, optimized SKW algorithm can be described by the repeated application of a unitary evolution operator $UU'$ on the Hilbert space ${\rm H} = {{\rm H}^C} \otimes {{\rm H}^S}$ of $n + 1$ dimensional hypercube. Each state in ${\rm H}$ can be described as $\left| {d,\vec x} \right\rangle $, where $d$ specifies the state of the coin and $\vec x$ specifies the position on the hypercube. The operators can be written as
\begin{equation}
U = S \cdot {C_0},
\end{equation}
\begin{equation}
U' = S \cdot C',
\end{equation}
where the shift operator is
 \begin{equation}
 S = \sum\limits_{x = 0}^{{2^{n + 1}} - 1} {\sum\limits_{d = 1}^{n + 1} {\left| {d,\vec x \oplus {{\vec e}_d}} \right\rangle } \left\langle {d,\vec x} \right|}.
 \end{equation}
It depends on the state of the coin to map a state $\left| {d,\vec x} \right\rangle $ onto the state $\left| {d,\vec x \oplus {{\vec e}_d}} \right\rangle $, where $\left| {{{\vec e}_d}} \right\rangle  = \left| {0 \cdots 010 \cdots 0} \right\rangle $ is a null vector except for a single 1 entry at the $d$th component. The coin operator can be written as
 \begin{equation}
C' = {C_0} \otimes I + ({C_1} - {C_0}) \otimes \left| 0 \right\rangle \left\langle 0 \right|.
 \end{equation}
It consists of two parts ${C_0}$ and ${C_1}$. ${C_0}$ acts on unmarked vertices, which is generally chosen as Grover operator. i.e.
 \begin{equation}
{C_0} = G = 2\left| {{S^C}} \right\rangle \left\langle {{S^C}} \right| - I,
 \end{equation}
where $\left| {{S^C}} \right\rangle  = \frac{1}{{\sqrt {n + 1} }}\sum\limits_{d = 1}^{n + 1} {\left| d \right\rangle } $ is the equal superposition over all $n + 1$ directions. ${C_1}$ acts on the marked vertex, which is generally chosen as $- I$.

    Optimized SKW algorithm is implemented as follows:

(1) Initialize the quantum computer to the equal superposition over all states, $\left| {{\psi _0}} \right\rangle  = \left| {{S^C}} \right\rangle  \otimes \left| {{S^S}} \right\rangle $. By orthogonal projection ${P_e}$, $\left| {{\psi _0}} \right\rangle $ is mapped to the basis vectors that Hamming weight of vertices are even to obtain the initial state $\left| {{\psi _0}^{(e)}} \right\rangle $.

(2) Apply the evolution operator $UU' = S \cdot \left( {{C_0} \otimes I} \right) \cdot S \cdot C'$ to the initial state $\left| {{\psi _0}^{(e)}} \right\rangle $ with $\frac{\pi }{4}\sqrt {{2^n}} $ times. $C'$ is used to flip the phase of the marked state, $C'\left| {{S^C}} \right\rangle  \otimes \left| x \right\rangle  = {( - 1)^{f(x)}}\left| {{S^C}} \right\rangle  \otimes \left| x \right\rangle $, when $\left| x \right\rangle  = \left| 0 \right\rangle $, $f(x) = 1$; when $\left| x \right\rangle  \ne \left| 0 \right\rangle $, $f(x) = 0$.

(3) Measure the quantum state. The probability of marked state$\left| {{S^C}} \right\rangle  \otimes \left| 0 \right\rangle $ is $1 - O(\frac{1}{{n + 1}})$.

The coin operator $C'$ changes the phase to mark state in the algorithm, taking on the function of an oracle. It iterates $\frac{\pi }{4}\sqrt {{2^n}} $ times under ideal conditions and the success rate is 1 approximately. Each iteration is equivalent to twice walks on the hypercube, containing four operations with two coin-flip operations and two transfer operations. $C'$ is only applied once, so each iteration only calls oracle once.

In order to establish and analyze the model of optimized SKW algorithm with phase errors, the geometric description of optimized SKW algorithm will be given in the following.

\section{Geometric description of optimized SKW algorithm}  
\label{sec:2}
In order to simplify the problem, let us first show that the random walk on the hypercube can be collapsed onto a random walk on the line. We use the method presented in ${[8]}$. Let ${P_{ij}}$ be the permutation operator defined in Theorem 1 of [8].

\textbf{Theorem 1}  $U$ and $U'$ are the unitary operators in Section 2 and ${P_{ij}}$ is the permutation operator. Then $UU'$ and ${P_{ij}}$ satisfy $P_{ij}^ + UU'{P_{ij}} = UU'$.

\textbf{Proof}: It can be obtained from Theorem 1 in ${[8]}$ that $P_{ij}^ + U{P_{ij}} = U$ and $P_{ij}^ + U'{P_{ij}} = U'$. We multiply these two equations by the left and right respectively and the theorem is proved.                                     ■

Since the initial state $\left| {{\psi _0}^{(e)}} \right\rangle $ is an eigenvector of ${P_{ij}}$ with eigenvalue 1 for all $i$ and $j$. According to Theorem 1, any intermediate state ${(UU')^t}\left| {{\psi _0}^{(e)}} \right\rangle $ must also be an eigenvector of eigenvalue 1 with respect to ${P_{ij}}$. Thus ${(UU')^t}$ preserves the symmetry of $\left| {{\psi _0}^{(e)}} \right\rangle $ with respect to bit swaps. Define $2n$ basis states, $\left| {R,0} \right\rangle ,\left| {L,1} \right\rangle ,\left| {R,1} \right\rangle  \cdots \left| {R,n - 1} \right\rangle ,\left| {L,n} \right\rangle $, that their specific forms are the same with equation (10) and (11) in [8].
Based on the above basis states, the coin operator and shift operator will be rewrote respectively by equation (13) and (12) in [8]. So, the unitary operators also become the form of (14) and (15) in [8].

As both $U$ and $U'$ are real matrices, $UU$ and $UU'$ are also real matrices. Their eigenvalues and eigenvectors are complex conjugate pairs. Before analyzing the eigenvalue spectrum of $UU'$, let's analyze the eigenvalue spectrum of $UU$ first.

The eigenvalue spectrum of $U$ is already given in ${[15]}$.
Since $UU\left| {{\upsilon _{\vec k}}} \right\rangle  = U({e^{ \pm i{\omega _k}}}\left| {{\upsilon _{\vec k}}} \right\rangle ) = {e^{ \pm i2{\omega _k}}}\left| {{\upsilon _{\vec k}}} \right\rangle $, we obtain the eigenvalue spectrum of $UU$,
\begin{equation}
\begin{array}{lll}
{e^{ \pm i2{\omega _k}}} = \cos 2{\omega _k} \pm i\sin 2{\omega _k} \\
\quad\quad\quad= 1 + \frac{{8k(k - n)}}{{{n^2}}} \pm i\frac{{4(n - 2k)}}{{{n^2}}}\sqrt {k(n - k)}.
\end{array}
\end{equation}

\textbf{Theorem 2} $U$ and $U'$ are the unitary operators in Section 2. Then there are two eigenvectors of $UU'$ with eigenvalues close to 1.

\textbf{proof}: Supposing the size of the database is ${2^{n - 1}}$ and $n$ is a multiple of 4. We construct two approximate eigenvectors of $UU'$, $\left| {{\psi _0}^{(e)}} \right\rangle $ and $\left| {{\psi _1}} \right\rangle $,
\begin{equation}
\begin{array}{lll}
\left| {{\psi _0}^{(e)}} \right\rangle  = \frac{1}{{\sqrt {{2^{n - 1}}} }}\left| {R,0} \right\rangle+ \frac{1}{{\sqrt {{2^{n - 1}}} }}\left| {L,n} \right\rangle \\
\quad\quad\quad\quad+ \sum\limits_{x = 1}^{\frac{n}{2} - 1} ( \sqrt {\frac{{\left( {\begin{array}{*{20}{c}}
{n - 1}\\
{2x}
\end{array}} \right)}}{{{2^{n - 1}}}}} \left| {R,2x} \right\rangle\nonumber \\
 \quad\quad\quad\quad+ \sqrt {\frac{{\left( {\begin{array}{*{20}{c}}
{n - 1}\\
{2x - 1}
\end{array}} \right)}}{{{2^{n - 1}}}}} \left| {L,2x} \right\rangle ),
\end{array}
\end{equation}

\begin{equation}
\begin{array}{lll}
\left| {{\psi _1}} \right\rangle  = \frac{1}{c}(\sum\limits_{x = 0}^{\frac{n}{4} - 1} {\frac{1}{{\sqrt {\left( {\begin{array}{*{20}{c}}
{n - 1}\\
{2x}
\end{array}} \right)} }}} \left| {R,2x} \right\rangle \\
 \quad\quad\quad\quad- \frac{1}{{\sqrt {\left( {\begin{array}{*{20}{c}}
{n - 1}\\
{2x + 1}
\end{array}} \right)} }}\left| {L,2x + 2} \right\rangle ).
\end{array}
\end{equation}

where $c = \sqrt {\sum\limits_{x = 0}^{\frac{n}{2} - 1} {\frac{1}{{\left( {\begin{array}{*{20}{c}}
{n - 1}\\
x
\end{array}} \right)}}} } $. We evaluate

\begin{equation}
UU'\left| {{\psi _0}^{(e)}} \right\rangle  = \left| {{\psi _0}^{(e)}} \right\rangle  - \frac{2}{{\sqrt {{2^{n - 1}}} }}(\frac{{2 - n}}{n}\left| {R,0} \right\rangle  + \frac{{2\sqrt {n - 1} }}{n}\left| {L,2} \right\rangle ),
\end{equation}
obtaining
\begin{equation}
\left\langle {{\psi _0}^{(e)}} \right|UU'\left| {{\psi _0}^{(e)}} \right\rangle  =1 - \frac{1}{{{2^{n - 2}}}}.
\end{equation}
So, when $n$ is large enough, $\left| {{\psi _0}^{(e)}} \right\rangle $ is an eigenvector of $UU'$ with eigenvalue close to 1. Then we evaluate
\begin{equation}
\begin{array}{lll}
UU'\left| {{\psi _1}} \right\rangle \\
 = \left| {{\psi _1}} \right\rangle  - \frac{1}{c}(1 - \frac{2}{n})\frac{1}{{\sqrt {\left( {\begin{array}{*{20}{c}}
{n - 1}\\
{\frac{n}{2} - 2}
\end{array}} \right)} }}(\left| {R,\frac{n}{2} - 2} \right\rangle  + \left| {L,\frac{n}{2} + 2} \right\rangle )\\
\quad - \frac{1}{c}\frac{2}{n}\frac{1}{{\sqrt {\left( {\begin{array}{*{20}{c}}
{n - 1}\\
{\frac{n}{2} - 1}
\end{array}} \right)} }}(\left| {R,\frac{n}{2}} \right\rangle  + \left| {L,\frac{n}{2}} \right\rangle )
\end{array}
\end{equation}
in the same way to obtain
\begin{equation}
\left\langle {{\psi _1}} \right|UU'\left| {{\psi _1}} \right\rangle  = 1 - \frac{1}{{{c^2}\left( {\begin{array}{*{20}{c}}
{n - 1}\\
{{\raise0.7ex\hbox{$n$} \!\mathord{\left/
 {\vphantom {n 2}}\right.\kern-\nulldelimiterspace}
\!\lower0.7ex\hbox{$2$}}}
\end{array}} \right)}}.
\end{equation}
Since $1 < {c^2} < 1 + \frac{2}{n}$, $\left| {{\psi _1}} \right\rangle $ is also an eigenvector of $UU'$ with eigenvalue close to 1. The theorem is proved.                                                                                                                     ■

Then according to Theorem 2,3 in ${[8]}$ and the eigenvalue spectrum of $UU$, it can be obtained that there are exactly two eigenvalues ${e^{i{{\omega '}_0}}},{e^{ - i{{\omega '}_0}}}$ of $UU'$ with their real parts bigger than
\begin{equation}
L = 1 - \frac{{8n}}{{3{{(n + 1)}^2}}}.
\end{equation}
Set the corresponding eigenvectors $\left| {{{\omega '}_0}} \right\rangle $ and $\left| { - {{\omega '}_0}} \right\rangle $, satisfying $\left| {{{\omega '}_0}} \right\rangle  = {\left| { - {{\omega '}_0}} \right\rangle ^*}$. According to the Theorem 4 in ${[8]}$, $\left| {{{\omega '}_0}} \right\rangle $ and $\left| { - {{\omega '}_0}} \right\rangle $ can be also proved to be approximately composed of the linear combination of $\left| {{\psi _0}^{(e)}} \right\rangle $ and $\left| {{\psi _1}} \right\rangle $,
\begin{equation}
\left| {{{\omega '}_0}} \right\rangle  \approx \frac{1}{{\sqrt 2 }}(\left| {{\psi _0}^{(e)}} \right\rangle  + i\left| {{\psi _1}} \right\rangle ),
\end{equation}

\begin{equation}
\left| { - {{\omega '}_0}} \right\rangle  \approx \frac{1}{{\sqrt 2 }}(\left| {{\psi _0}^{(e)}} \right\rangle  - i\left| {{\psi _1}} \right\rangle ).
\end{equation}

Finally, we evaluate the range of ${\omega '_0}$. First we evaluate to obtain
\begin{equation}
\left\langle {{\psi _1}} \right|UU'\left| {{\psi _0}^{(e)}} \right\rangle  = \left\langle {{{\psi _1}}}
 \mathrel{\left | {\vphantom {{{\psi _1}} {{\psi _0}^{(e)}}}}
 \right. \kern-\nulldelimiterspace}
 {{{\psi _0}^{(e)}}} \right\rangle  + \frac{1}{c}\frac{2}{{\sqrt {{2^n}} }},
\end{equation}
\begin{equation}
\left\langle {{\psi _0}^{(e)}} \right|UU'\left| {{\psi _1}} \right\rangle  = \left\langle {{{\psi _0}^{(e)}}}
 \mathrel{\left | {\vphantom {{{\psi _0}^{(e)}} {{\psi _1}}}}
 \right. \kern-\nulldelimiterspace}
 {{{\psi _1}}} \right\rangle  - \frac{1}{c}\frac{2}{{\sqrt {{2^n}} }}.
\end{equation}
Supposing $\left| \alpha  \right\rangle  = \frac{1}{{\sqrt 2 }}(\left| {{\psi _0}^{(e)}} \right\rangle  + i\left| {{\psi _1}} \right\rangle )$, according to the Theorem 5 in ${[8]}$,
\begin{equation}
\begin{array}{lll}
|{e^{i{{\omega '}_0}}} - \left\langle \alpha  \right|U'\left| \alpha  \right\rangle | \\
\le 2(\frac{{3{{(n + 1)}^2}}}{{n \cdot {2^{n + 4}}}} + \frac{{3{{(n + 1)}^2}}}{{16{c^2} \cdot n \cdot \left( {\begin{array}{*{20}{c}}
n\\
{\frac{n}{2} + 1}
\end{array}} \right)}}) = O(\frac{{{n^{{\raise0.7ex\hbox{$3$} \!\mathord{\left/
 {\vphantom {3 2}}\right.\kern-\nulldelimiterspace}
\!\lower0.7ex\hbox{$2$}}}}}}{{{2^n}}})
\end{array}
\end{equation}
is obtained. The imaginary part is
 \begin{equation}
\begin{array}{lll}
{\mathop{\rm Im}\nolimits} \left\langle \alpha  \right|U'\left| \alpha  \right\rangle  = \frac{1}{2}(\left\langle {{\psi _0}^{(e)}} \right|UU'\left| {{\psi _1}} \right\rangle  - \left\langle {{\psi _1}} \right|UU'\left| {{\psi _0}^{(e)}} \right\rangle )\\
 \quad\quad\quad\quad\quad\quad=  - \frac{1}{c}\frac{2}{{\sqrt {{2^n}} }}.
\end{array}
\end{equation}
On the other hand, because of
\begin{eqnarray}
|\sin {\omega '_0} - {\mathop{\rm Im}\nolimits} \left\langle \alpha  \right|U'\left| \alpha  \right\rangle | = |{\mathop{\rm Im}\nolimits} ({e^{i{{\omega '}_0}}} - \left\langle \alpha  \right|U'\left| \alpha  \right\rangle )| \nonumber\\\le |{e^{i{{\omega '}_0}}} - \left\langle \alpha  \right|U'\left| \alpha  \right\rangle | \le O(\frac{{{n^{{\raise0.7ex\hbox{$3$} \!\mathord{\left/
 {\vphantom {3 2}}\right.\kern-\nulldelimiterspace}
\!\lower0.7ex\hbox{$2$}}}}}}{{{2^n}}})
\end{eqnarray}
so,
 \begin{equation}
 |\sin {\omega '_0} + \frac{2}{{c \cdot \sqrt {{2^n}} }}| \le O(\frac{{{n^{{\raise0.7ex\hbox{$3$} \!\mathord{\left/
 {\vphantom {3 2}}\right.\kern-\nulldelimiterspace}
\!\lower0.7ex\hbox{$2$}}}}}}{{{2^n}}}).
 \end{equation}
When ${\omega '_0}$ tends to 0, there will be $\sin {\omega '_0} = {\omega '_0} + O({\omega '_0}^3)$. So
  \begin{equation}
 - \frac{2}{{c \cdot \sqrt {{2^n}} }} - O(\frac{{{n^{{\raise0.7ex\hbox{$3$} \!\mathord{\left/
 {\vphantom {3 2}}\right.\kern-\nulldelimiterspace}
\!\lower0.7ex\hbox{$2$}}}}}}{{{2^n}}}) \le {\omega '_0} \le  - \frac{2}{{c \cdot \sqrt {{2^n}} }} + O(\frac{{{n^{{\raise0.7ex\hbox{$3$} \!\mathord{\left/
 {\vphantom {3 2}}\right.\kern-\nulldelimiterspace}
\!\lower0.7ex\hbox{$2$}}}}}}{{{2^n}}}).
 \end{equation}

At last this algorithm can be geometrically described through the above conclusions. Starting with the initial state $\left| {{\psi _0}^{(e)}} \right\rangle $, it successively iterates $t$ times,
\begin{eqnarray}
{\left( {UU'} \right)^t}\left| {{\psi _0}^{(e)}} \right\rangle  = \frac{1}{{\sqrt 2 }}({e^{i{{\omega '}_0}t}}\left| {{{\omega '}_0}} \right\rangle  + {e^{ - i{{\omega '}_0}t}}\left| { - {{\omega '}_0}} \right\rangle )\nonumber\\= \cos {\omega '_0}t\left| {{\psi _0}^{(e)}} \right\rangle  - \sin {\omega '_0}t\left| {{\psi _1}} \right\rangle.
\end{eqnarray}
Each iteration is equivalent to a rotation of angle $|{\omega '_0}|$ on the plane composed of the initial state $\left| {{\psi _0}^{(e)}} \right\rangle  = \frac{1}{{\sqrt 2 }}(\left| {{{\omega '}_0}} \right\rangle  + \left| { - {{\omega '}_0}} \right\rangle )$ and $\left| {{\psi _1}} \right\rangle  = \frac{i}{{\sqrt 2 }}(\left| {{{\omega '}_0}} \right\rangle  - \left| { - {{\omega '}_0}} \right\rangle )$ (shown in Fig. 1). Because of $\left\langle {{{\psi _0}^{(e)}}}
 \mathrel{\left | {\vphantom {{{\psi _0}^{(e)}} {{\psi _1}}}}
 \right. \kern-\nulldelimiterspace}
 {{{\psi _1}}} \right\rangle  = 0$, the required iteration times is $\frac{\pi }{{2|{{\omega '}_0}|}} = \frac{\pi }{4}\sqrt {{2^n}} $. After obtaining $\left| {{\psi _1}} \right\rangle $, the marked state $\left| {R,0} \right\rangle $ can be measured with probability $p = {\left| {\left\langle {{R,0}}
 \mathrel{\left | {\vphantom {{R,0} {{\psi _1}}}}
 \right. \kern-\nulldelimiterspace}
 {{{\psi _1}}} \right\rangle } \right|^2} = \frac{1}{{{c^2}}}$. So the success rate of the algorithm is $1 - O(\frac{1}{{n + 1}})$. Both the number of iterations and the success rate meet the conclusions in [9]. The geometric description of optimized SKW algorithm has been presented.
 \begin{figure}[htb]
  \resizebox{9.3cm}{7cm}{
  \includegraphics{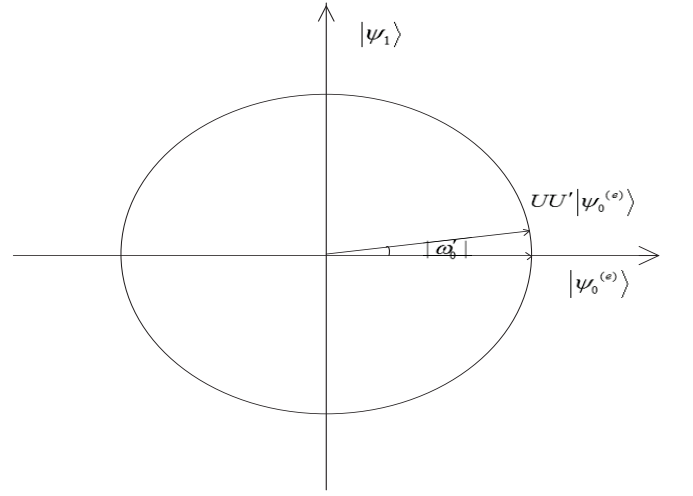}}

\caption{Each iteration of the algorithm $UU'$ can be viewed as a rotation on the plane composed of the initial state $\left| {{\psi _0}^{(e)}} \right\rangle $ and the final state $\left| {{\psi _1}} \right\rangle $. The rotation angle is $|{\omega '_0}|$ }
\label{fig:1}       
\end{figure}

\section{Model of the algorithm with phase errors}  
\label{sec:3}
\subsection{Build the model}  
\label{sec:4}
The model of the algorithm with phase errors can be built after the geometric description of the original algorithm. Let the database size be ${2^{n - 1}}$. As the Grover operator imperfections in phase inversions of the algorithm are systematic, the explicit form of the inaccurate operator can be written as
 \begin{equation}
\widetilde {{C_0}} = G = (1 - {e^{i\theta }})\left| {{S^C}} \right\rangle \left\langle {{S^C}} \right| - I,
 \end{equation}
where $\theta  = \pi  + \delta $ and $\delta $ is a constant error. When $\delta  = 0$, it recovers the original operator. Then the coin operator with errors can be written as
\begin{equation}
\widetilde {C'} = \widetilde {{C_0}} \otimes I + ({C_1} - \widetilde {{C_0}}) \otimes \left| 0 \right\rangle \left\langle 0 \right|.
 \end{equation}
The evolution operator becomes
\begin{equation}
\widetilde U = S \cdot \widetilde {{C_0}},
 \end{equation}
 \begin{equation}
\widetilde {U'} = S \cdot \widetilde {C'}.
 \end{equation}
Each iteration of the algorithm becomes $\widetilde U\widetilde {U'}$.

We analyze the eigenvalue spectrum of $\widetilde U\widetilde U$ first. The eigenvalue spectrum of $\widetilde U$ is already given by ${[14]}$ with
 \begin{equation}
{e^{i\widetilde {{\omega _k}}}} = \frac{1}{2}(1 - {e^{i\theta }}) - (1 - {e^{i\theta }})\frac{k}{n} \mp \frac{1}{2}\sqrt {{{(1 - {e^{i\theta }})}^2}\frac{{{{(n - 2k)}^2}}}{{{n^2}}} + 4{e^{i\theta }}},
 \end{equation}
$k \ne 0,k \ne n$.When $k = 0$, ${e^{i\widetilde {{\omega _0}}}} =  - {e^{i\theta }} = \cos \delta  + i\sin \delta $; $k = n$, ${e^{i\widetilde {{\omega _n}}}}= {e^{i\theta }}$. The introduction of errors lead to an $\delta $ angle deviation between ${e^{i\widetilde {{\omega _0}}}}$ and ${e^{i{\omega _0}}} = 1$ on the polar coordinate plane (see fig. 4.1 in ${[17]}$). The eigenvalue spectrum of $\widetilde U\widetilde U$ is ${e^{2i\widetilde {{\omega _k}}}}$, when $k = 0$, ${e^{2i\widetilde {{\omega _0}}}} = {e^{2i\theta }} = \cos 2\delta  + i\sin 2\delta $. So, there is an $2\delta $ angle deviation between ${e^{2i\widetilde {{\omega _0}}}}$ and ${e^{2i{\omega _0}}}$ on the polar coordinate plane. Eigenvalues are no longer complex conjugate pairs.

For the eigenvalue of $\widetilde U\widetilde {U'}$,  it only needs to analyze ${e^{i\widetilde {{{\omega '}_0}}}}$, ${e^{i\widetilde {{{\omega '}_1}}}}$, which originally correspond to the eigenvectors $\left| {{{\omega '}_0}} \right\rangle $ and $\left| { - {{\omega '}_0}} \right\rangle $ , because they determine the initial state and the final state of the algorithm. These two eigenvalues are conjugate pairs originally. But the introduction of errors lead to them being no longer conjugate. When   $\delta $ increases, the difference of these two eigenvalues gradually increased. ${e^{i\widetilde {{{\omega '}_0}}}}$ and ${e^{i\widetilde {{{\omega '}_1}}}}$ will gradually approach 1 and ${e^{i2\theta }}$ respectively according to ${[14]}$. If $\delta $ is very small, they are still in the vicinity of 1.

According to Theorem 1 and ${[14]}$, the random walk which introduces phase errors can also be collapsed onto a random walk on the line. To redefine the operator, where
 \begin{equation}
\widetilde {U'} = \widetilde U - (1 - {e^{i\theta }})\left| {L,1} \right\rangle \left\langle {R,0} \right|,
 \end{equation}
 the specific form of $\widetilde U$ has been redefined in ${[14]}$.

Expanding eigenvalues of $\widetilde U\widetilde U$ to the second order with respect to $\delta $ and using the Theorem 2 and Theorem 3 in ${[8]}$, it is can be proved that there are exactly two eigenvalues of $\widetilde U\widetilde {U'}$ with their real parts $\widetilde L$ bigger than
\begin{equation}
\begin{array}{lll}
1 - \frac{1}{3}\left[ {(\frac{8}{n} - \frac{8}{{{n^2}}}) - (2 - \frac{4}{n})\frac{{\sqrt {n - 1} }}{n}\delta  + 3(\frac{1}{3} + \frac{2}{{{n^2}}} - \frac{2}{n}){\delta ^2}} \right].
\end{array}
 \end{equation}
When $\delta  = 0$ and $n$ is replaced by $n + 1$, it is equal to equation (13). Then we evaluate
 \begin{equation}
 \begin{array}{lll}
\widetilde U\widetilde {U'}\left| {{\psi _0}^{(e)}} \right\rangle  = {e^{2i\theta }}\left| {{\psi _0}^{(e)}} \right\rangle  - \frac{{1 - {e^{i\theta }}}}{{\sqrt {{2^{n - 1}}} }}(\frac{{1 - {e^{i\theta }} - n}}{n}\left| {R,0} \right\rangle  \\
\quad\quad\quad\quad\quad\quad+ \frac{{(1 - {e^{i\theta }})\sqrt {n - 1} }}{n}\left| {L,2} \right\rangle ),
 \end{array}
 \end{equation}
so
 \begin{equation}
  \begin{array}{lll}
\left\langle {{\psi _0}^{(e)}} \right|\widetilde U\widetilde {U'}\left| {{\psi _0}^{(e)}} \right\rangle  = {e^{2i\theta }} - \frac{{({e^{i\theta }} - 1){e^{i\theta }}}}{{{2^{n - 1}}}} \\
\quad\quad\quad\quad\quad\quad\quad\quad\quad= {e^{2i\delta }} - \frac{{({e^{i\delta }} + 1){e^{i\delta }}}}{{{2^{n - 1}}}}.
 \end{array}
  \end{equation}
In the same way
\begin{equation}
 \begin{array}{lll}
\widetilde U\widetilde {U'}\left| {{\psi _1}} \right\rangle  \\
= \left| {{\psi _1}} \right\rangle  - \frac{{{{(1 - {e^{i\theta }})}^2}}}{{c \cdot 4}}(1 - \frac{2}{n})\frac{1}{{\sqrt {\left( {\begin{array}{*{20}{c}}
{n - 1}\\
{\frac{n}{2} - 2}
\end{array}} \right)} }}(\left| {R,\frac{n}{2} - 2} \right\rangle  + \left| {L,\frac{n}{2} + 2} \right\rangle )\\
- \frac{1}{c}(\frac{{1 - {e^{i2\theta }}}}{4} + \frac{{{{(1 - {e^{i\theta }})}^2}}}{{2n}})\frac{1}{{\sqrt {\left( {\begin{array}{*{20}{c}}
{n - 1}\\
{\frac{n}{2} - 1}
\end{array}} \right)} }}(\left| {R,\frac{n}{2}} \right\rangle  + \left| {L,\frac{n}{2}} \right\rangle ),
 \end{array}
 \end{equation}
so
 \begin{equation}
\left\langle {{\psi _1}} \right|\widetilde U\widetilde {U'}\left| {{\psi _1}} \right\rangle  = 1 - \frac{{({e^{i\theta }} - 1){e^{i\theta }}}}{{2{c^2}\left( {\begin{array}{*{20}{c}}
{n - 1}\\
{{\raise0.7ex\hbox{$n$} \!\mathord{\left/
 {\vphantom {n 2}}\right.\kern-\nulldelimiterspace}
\!\lower0.7ex\hbox{$2$}}}
\end{array}} \right)}} = 1 - \frac{{({e^{i\delta }} + 1){e^{i\delta }}}}{{2{c^2}\left( {\begin{array}{*{20}{c}}
{n - 1}\\
{{\raise0.7ex\hbox{$n$} \!\mathord{\left/
 {\vphantom {n 2}}\right.\kern-\nulldelimiterspace}
\!\lower0.7ex\hbox{$2$}}}
\end{array}} \right)}}.
 \end{equation}

From the above equations (32) and (34), when $n$ is big enough, $\left| {{\psi _0}^{(e)}} \right\rangle $ is an eigenvector of $\widetilde U\widetilde {U'}$ with eigenvalue close to ${e^{2i\delta }}$ and $\left| {{\psi _1}} \right\rangle $ is an eigenvector of $\left| {{\psi _1}} \right\rangle $ with eigenvalue close to 1. Because ${e^{2i\delta }} = \cos 2\delta {\rm{ + isin2}}\delta $, its real part can be expanded by Taylor expansion $\cos 2\delta  = 1 - 2{\delta ^2}$. Based on the analysis above, when the error is small, the algorithm can still be viewed as rotation on the two-dimensional plane composed of (7) and (8).

\subsection{analysis of the algorithm}
\label{sec:5}
With the error increasing gradually, $\left| {\widetilde {{{\omega '}_0}}} \right\rangle $ and $\left| {\widetilde {{{\omega '}_1}}} \right\rangle $ are no longer conjugate. They can't be expressed as the linear combination of (14) and (15). So the algorithm can't rotate from $\left| {{\psi _0}^{(e)}} \right\rangle $ to $\left| {{\psi _1}} \right\rangle $ completely, and the probability to obtain $\left| {{\psi _1}} \right\rangle $ becomes small finally.

Expanding $\left| {{\psi _0}^{(e)}} \right\rangle $ and $\left| {{\psi _1}} \right\rangle $ with $\left| {\widetilde {{{\omega '}_0}}} \right\rangle $ and $\left| {\widetilde {{{\omega '}_1}}} \right\rangle $ respectively, we obtain
 \begin{equation}
\left| {{\psi _0}^{(e)}} \right\rangle  = {a_0}\left| {\widetilde {{{\omega '}_0}}} \right\rangle  + {a_1}\left| {\widetilde {{{\omega '}_1}}} \right\rangle  + \sqrt {1 - a_0^2 - a_1^2} \left| {{\gamma _0}} \right\rangle,
 \end{equation}
 \begin{equation}
\left| {{\psi _1}} \right\rangle  = {b_0}\left| {\widetilde {{{\omega '}_0}}} \right\rangle  + {b_1}\left| {\widetilde {{{\omega '}_1}}} \right\rangle  + \sqrt {1 - b_0^2 - b_1^2} \left| {{\gamma _1}} \right\rangle,
 \end{equation}
where $\left| {{\gamma _0}} \right\rangle $ and $\left| {{\gamma _1}} \right\rangle $ are the normalized vectors orthogonal to $\left| {\widetilde {{{\omega '}_0}}} \right\rangle $ and $\left| {\widetilde {{{\omega '}_1}}} \right\rangle $. ${a_0},{a_1},{b_0},{b_1}$ are the complex numbers, decided by $\delta $ and $n$. Evaluating
\begin{equation}
\begin{array}{lll}
1 - \frac{1}{{{2^{n - 2}}}} - {\delta ^2}(2 - \frac{5}{{{2^n}}})\\
 = {\mathop{\rm Re}\nolimits} (\left\langle {{\psi _0}^{(e)}} \right|\widetilde U\widetilde {U'}\left| {{\psi _0}^{(e)}} \right\rangle )\\
  = \sum\limits_j {\cos \widetilde {{\omega _j}}} \left| {{{\left\langle {{\widetilde {{\omega _j}}}}
 \mathrel{\left | {\vphantom {{\widetilde {{\omega _j}}} {{\psi _0}^{(e)}}}}
 \right. \kern-\nulldelimiterspace}
 {{{\psi _0}^{(e)}}} \right\rangle }^2}} \right|\,\\
  = a_0^2\cos \widetilde {{\omega _0}} + a_1^2\cos \widetilde {{\omega _1}} + \sum\limits_{j \ne 0,1} {\cos \widetilde {{\omega _j}}} \left| {{{\left\langle {{\widetilde {{\omega _j}}}}
 \mathrel{\left | {\vphantom {{\widetilde {{\omega _j}}} {{\psi _0}^{(e)}}}}
 \right. \kern-\nulldelimiterspace}
 {{{\psi _0}^{(e)}}} \right\rangle }^2}} \right|\,\,\\ < a_0^2 + a_1^2 + \widetilde L(1 - a_0^2 - a_1^2)\quad,
\end{array}
\end{equation}
we obtain
 \begin{equation}
a_0^2 + a_1^2 > 1 - \frac{1}{{1 - {\widetilde L}}}[\frac{1}{{{2^{n - 2}}}} + (2 - \frac{5}{{{2^n}}}){\delta ^2}].
 \end{equation}
In the same way, evaluating ${\mathop{\rm Re}\nolimits} \left\langle {{\psi _1}} \right|\widetilde U\widetilde {U'}\left| {{\psi _1}} \right\rangle $, we obtain
\begin{equation}
b_0^2 + b_1^2 > 1 - \frac{1}{{1 - {\widetilde L}}}(\frac{{4 - 5{\delta ^2}}}{{4{c^2}\left( {\begin{array}{*{20}{c}}
{n - 1}\\
{{\raise0.7ex\hbox{$n$} \!\mathord{\left/
 {\vphantom {n 2}}\right.\kern-\nulldelimiterspace}
\!\lower0.7ex\hbox{$2$}}}
\end{array}} \right)}}).
 \end{equation}

Now the algorithm in the presence of phase errors can be geometrically described through the above conclusions. Applying the operator on the initial state $\left| {{\psi _0}^{(e)}} \right\rangle $ successively with $t$ times, we get
\begin{equation}
\begin{array}{lll}
{(\widetilde U\widetilde {U'})^t}\left| {{\psi _0}^{(e)}} \right\rangle\\
  = {a_0}{e^{i\widetilde {{{\omega '}_0}}t}}\left| {\widetilde {{{\omega '}_0}}} \right\rangle  + {a_1}{e^{i\widetilde {{{\omega '}_1}}t}}\left| {\widetilde {{{\omega '}_1}}} \right\rangle  + \sqrt {1 - a_0^2 - a_1^2} \left| {\gamma _0^t} \right\rangle\\
  = \frac{{{a_0}{b_1}{e^{i\widetilde {{{\omega '}_0}}t}} - {a_1}{b_0}{e^{i\widetilde {{{\omega '}_1}}t}}}}{{{a_0}{b_1} - {a_1}{b_0}}}(\left| {{\psi _0}^{(e)}} \right\rangle  - \sqrt {{\varepsilon _0}} \left| {{\gamma _0}} \right\rangle )\\
   \quad\quad- \frac{{{a_0}{a_1}({e^{i\widetilde {{{\omega '}_0}}t}} - {e^{i\widetilde {{{\omega '}_1}}t}})}}{{{a_0}{b_1} - {a_1}{b_0}}}(\left| {{\psi _1}} \right\rangle  - \sqrt {{\varepsilon _1}} \left| {{\gamma _1}} \right\rangle ) + \sqrt {{\varepsilon _0}} \left| {\gamma _0^t} \right\rangle,
\end{array}
\end{equation}
where $\left| {\gamma _0^t} \right\rangle  = {(\widetilde U\widetilde {U'})^t}\left| {{\gamma _0}} \right\rangle $, $\sqrt {1 - a_0^2 - a_1^2}  = \sqrt {{\varepsilon _0}} $ and $\sqrt {1 - b_0^2 - b_1^2}  = \sqrt {{\varepsilon _1}} $. So, the amplitude of $\left| {{\psi _1}} \right\rangle $ is
\begin{equation}
w = \frac{{{a_0}{a_1}({e^{i\widetilde {{{\omega '}_0}}t}} - {e^{i\widetilde {{{\omega '}_1}}t}})}}{{{a_0}{b_1} - {a_1}{b_0}}},
 \end{equation}
decided by eigenvalues ${e^{i\widetilde {{{\omega '}_0}}}}$, ${e^{i\widetilde {{{\omega '}_1}}}}$, iteration times $t$ and coefficients ${a_0},{a_1},{b_0},{b_1}$.

When $\delta {\rm{ = }}0$, there are ${e^{i\widetilde {{{\omega '}_0}}}} = {e^{i{{\omega '}_0}}}$, ${e^{i\widetilde {{{\omega '}_1}}}} = {e^{ - i{{\omega '}_0}}}$, ${a_0}{\rm{ = }}{a_1}{\rm{ = }}\frac{1}{{\sqrt 2 }}$, ${b_0}{\rm{ = }} - {b_1} = \frac{i}{{\sqrt 2 }}$ and ${\varepsilon _0},{\varepsilon _1} \to 0$. So $w = \sin {\omega _0}t$, it means that $\left| {{\psi _0}^{(e)}} \right\rangle $ can completely rotate to $\left| {{\psi _1}} \right\rangle $ finally. The probability to obtain $\left| {{\psi _1}} \right\rangle $ is nearly 1. When $\delta  \ne 0$, ${\varepsilon _0},{\varepsilon _1}$ are no longer close to 0 and there is another state other than $\left| {{\psi _1}} \right\rangle $ at last. So the probability to obtain $\left| {{\psi _1}} \right\rangle $ will decrease. On the other hand, $\left\langle {{{\psi _1}}}
 \mathrel{\left | {\vphantom {{{\psi _1}} {{\psi _0}^{(e)}}}}
 \right. \kern-\nulldelimiterspace}
 {{{\psi _0}^{(e)}}} \right\rangle  = 0$, $\left| {{\psi _0}^{(e)}} \right\rangle $ needs to rotate angle of $\frac{\pi }{2}$ on the plane composed of $\left| {{\psi _0}^{(e)}} \right\rangle $ and $\left| {{\psi _1}} \right\rangle $ in theory. It is equivalent to making $\left| {\widetilde {{{\omega '}_0}}t - \widetilde {{{\omega '}_1}}t} \right| = \pi $ at last. When $\delta $ increases, the difference between eigenvalues $\widetilde {{{\omega '}_0}} - \widetilde {{{\omega '}_1}}$ will increase. So, the number of iterations $t$ to reach maximum $w$ will decrease. There is ${e^{it}}$ in the equation (41), so it is a periodic function. Therefore when the algorithm is in the presence of phase errors, its success rate will still periodically change with the number of iterations.

The above analysis shows that, after the introduction of errors, the success rate of the algorithm decreases and the required number of iterations decreases too. The success rate still periodically changes with the number of iterations.

\section{Numerical simulations and robustness comparison}  
\label{sec:6}
Through the above analysis and proof, the relationship between the size of database, the phase error, the number of iterations and the amplitude of the marked state is obtained. We also simply analyze the effects of the phase error on the algorithm. In this section, we will respectively analyze the impact of phase errors on the success rate and the number of iterations through numerical simulations. Finally, the quantitative relationship between the size of database, the phase error, the number of iterations and the success rate is given.

Firstly, we simulate the relation between the number of iterations and the success rate when $n = 8$. We compare the two cases, where $\delta  = 0$ and $\delta  = 0.2$, as shown in Fig. 2.
\begin{figure}[htb]
  \resizebox{9.3cm}{7cm}{\includegraphics{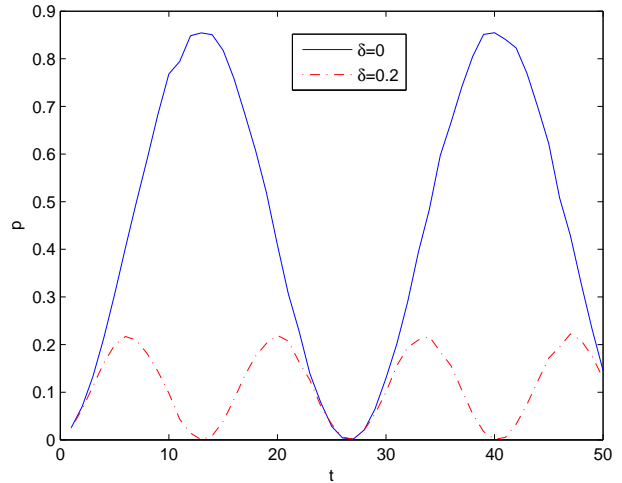}}

\caption{The relation between the success rate of the algorithm $p$ and the number of iterations $t$ when $n = 8$. The solid line represents $\delta  = 0.2$ and the dotted line represents $\delta  = 0$ }
\label{fig:2}       
\end{figure}

The solid line indicates that the algorithm without errors iterates 13 times to reach the maximum probability. The dotted line represents that it needs 6 iteration times to reach the maximum probability when the error is 0.2. But this time the maximum probability decreases compared to the former. Both the success rates periodically change with the number of iterations. Simulation results are consistent with the theoretical analysis in Section 4.2.

Then we study the relationship between the maximum success rate of the algorithm, the size of database and the phase error. Each curve represents different phase errors, with $\delta  = 0,\delta  = 0.01,\delta  = 0.001$ and $\delta  = 0.0001$, as shown in Fig. 3.
\begin{figure}[htb]
 \resizebox{9.3cm}{7cm}{\includegraphics{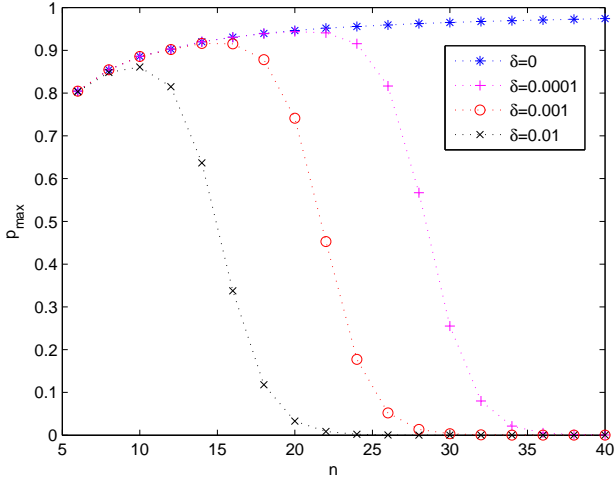}}

\caption{The maximum success rate of the algorithm ${p_{\max }}$ vs the scale of the database $n$ with different phase errors. The asterisks line for $\delta  = 0$, the ‘+’ line for $\delta  = 0.0001$ , the circles line for $\delta  = 0.001$ , the ‘x’ line for $\delta  = 0.01$ }
\label{fig:3}       
\end{figure}

It is shown in Fig. 3 that the maximum success rate increases with the increment of the database size first. When it reaches a threshold, the maximum success rate decreases exponentially. And the larger the error is, the smaller the database scale will be when the maximum success rate begins to decrease. With curve fitting, we obtain the expression about the maximum success rate ${p_{\max }}$, the size of database ${2^n}$ and the phase error $\delta $,
\begin{equation}
{p_{\max }} = \frac{{3.8{p_0}}}{{3.8 + {\delta ^2}{2^n}}}.
 \end{equation}
${p_0}$ is the success rate of the algorithm without errors. It can be obtained from Section 4
\begin{equation}
{p_0} = \frac{1}{{\sum\limits_{x = 0}^{{\raise0.7ex\hbox{$n$} \!\mathord{\left/
 {\vphantom {n 2}}\right.\kern-\nulldelimiterspace}
\!\lower0.7ex\hbox{$2$}} - 1} {\frac{1}{{\left( {\begin{array}{*{20}{c}}
{n - 1}\\
x
\end{array}} \right)}}} }} = \frac{1}{{{c^2}}} \to 1.
 \end{equation}
Combining the above two equations (42) and (43), we obtain
\begin{equation}
{p_{\max }} = \frac{{19}}{{{c^2}(19 + 5{\delta ^2} \cdot {2^n})}}.
 \end{equation}

It can be known from equation (44) that before the search in order to ensure the success rate of the algorithm, we need to predict the size of the phase error to limit the size of the database. For example, if the requirements of the success rate is not less than 0.5, the database size can't be more than $\frac{{3.8}}{{{\delta ^2}}}$.

Then we study the relationship between the iteration times required to reach the maximum success rate, the size of the database and the phase error. The corresponding simulation results are shown in Fig. 4.
\begin{figure}[htb]
 \resizebox{9.3cm}{7cm}{\includegraphics{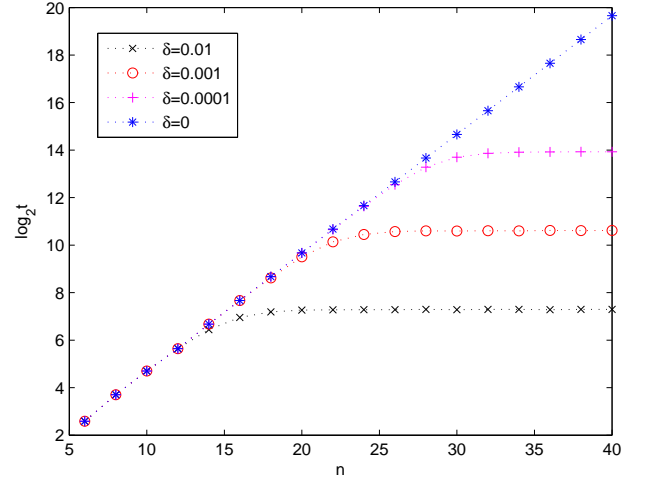}}

\caption{The scale of the required iteration times ${\log _2}{t}$ vs the scale of the database $n$ with different phase errors. The asterisks line for $\delta  = 0$, the ‘+’ line for $\delta  = 0.0001$ , the circles line for $\delta  = 0.001$ , the ‘x’ line for $\delta  = 0.01$ }
\label{fig:4}       
\end{figure}

It is shown in Fig. 4 that the required iteration times ${t_o}$ increases with the increment of the database first. Then it remains the same. When the required iteration times begin to remain the same, the larger the error is, the smaller the database scale will be. With curve fitting, we obtain the expression about iteration times ${t_o}$, the size of the database ${2^n}$ and the phase error $\delta $,
\begin{equation}
{t_o} = \frac{\pi }{{\sqrt {{\raise0.7ex\hbox{${16}$} \!\mathord{\left/
 {\vphantom {{16} {{2^n}}}}\right.\kern-\nulldelimiterspace}
\!\lower0.7ex\hbox{${{2^n}}$}} + 4{\delta ^2}} }}.
 \end{equation}

From the above analysis, when it is in the presence of phase errors, as long as the database scale does not exceed a critical value, the maximum success rate of the algorithm and the required number of iterations are the same with the original algorithm respectively. It suggests that this algorithm has certain tolerance to this kind of errors. Comparing Fig. 3 with Fig. 4 for the same error, we find that the corresponding database scale when ${p_{\max }}$ begins to decrease is smaller than that when ${t_o}$ begins to remain the same. It is shown ${p_{\max }}$ is more sensitive to the influence of the error. So, we take the value of ${n}$ when ${p_{\max }}$ begins to decrease as the critical value. ${p_{\max }}$ is already obtained. When ${p_{\max }}$ begins to decrease, we obtain the corresponding $\left( {\delta,{\rm{n}} } \right)$ by numerical simulations. The expression about the database scale ${n}$ and the phase error $\delta $ is
\begin{equation}
{\rm{n}} = 1.806 \cdot \log _2^{\frac{{0.4642}}{\delta }}.
 \end{equation}
So, when the algorithm is in the presence of errors, it is almost unaffected as long as the database scale satisfies ${\rm{n}} \le 1.806 \cdot \log _2^{\frac{{0.4642}}{\delta }}$.

Now we respectively obtain the maximum success rate ${p_{\max }}$ and the required iteration times ${t_o}$ expressions about the phase error $\delta $ and the database size ${2^n}$. From the analysis in Section 4.2, the probability to obtain marked state $p$ is a periodic function, which can also be confirmed from Fig. 2. So, making use of equation (44) and (45), we obtain an expression of the database size ${2^n}$, the phase error $\delta $, the number of iterations $t$ and the success rate $p$,
\begin{equation}
p = \frac{{19}}{{{c^2}(19 + 5{\delta ^2} \cdot {2^n})}}{\sin ^2}(\sqrt {\frac{4}{{{2^n}}} + {\delta ^2}}  \cdot t).
 \end{equation}

According to equation (47), before the search it can depend on the requirement of the success rate to weigh the number of iterations according to the given sized database and the predicted error. It will improve the algorithm efficiency. It is shown in Fig. 5 when $n = 8$. When $\delta  = 0$ or $\delta  = 0.2$, it is observed exactly the same as that in Fig. 2.
\begin{figure}[htb]
 \resizebox{9.3cm}{7cm}{\includegraphics{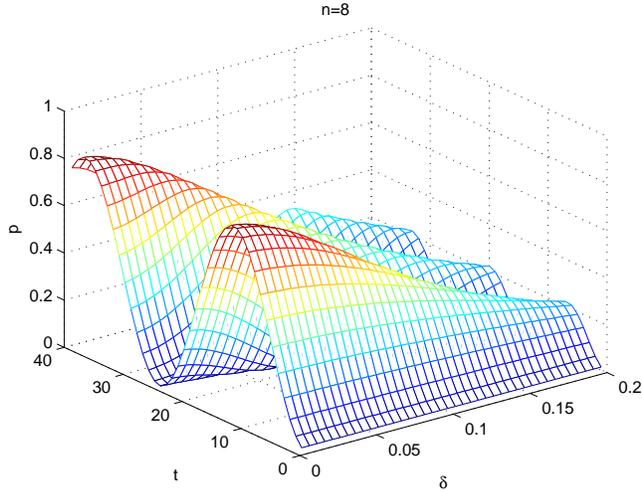}}

\caption{The three-dimensional graph of equation (47) when $n = 8$. The x-axis is the number of iterations $t$ . The y-axis is the phase error $\delta $  and the z-axis is the success rate of the algorithm $p$ }
\label{fig:5}       
\end{figure}

Finally, we make a simple comparison between optimized SKW algorithm and Grover's algorithm of robustness to the phase error. The definition of probability gap is the probability difference between the marked state and the maximum rest state$^{[16]}$. It is defined as
\begin{equation}
\Delta p = {p_{out}} - \max ({p_{2,}}{p_3} \cdots {p_{N - 1}}).
 \end{equation}
In practice, as long as there is sufficient probability gap, the probability of finding the marked state is much higher than other states at the end of the algorithm. It is still successful. If the algorithmic probability gap varies little before and after the presence of errors, its robustness is strong. So, we can measure the difference between their probability gaps before and after the presence of errors to compare robustness.

According to ${[17]}$, the expression of the database size, the phase error and the maximum success rate is given in Grover's algorithm,
\begin{equation}
{p_{\max }} = \frac{4}{{4 + {\delta ^2}{2^n}}}.
 \end{equation}
So, the probability gap of Grover's algorithm is
\begin{equation}
\Delta {p_2} = {p_{\max }} - \frac{{1 - {p_{\max }}}}{{{2^n} - 1}}.
 \end{equation}

Firstly, we simulate the probability gap of optimized SKW algorithm $\Delta{{p}_{1}}$ with different phase errors. It is shown in Fig. 6. Then we simulate the difference between the probability gaps $\Delta{{p}_{1}}-\Delta{{p}_{2}}$, shown in Fig. 7.

\begin{figure}[htb]
 \resizebox{9.3cm}{7cm}{\includegraphics{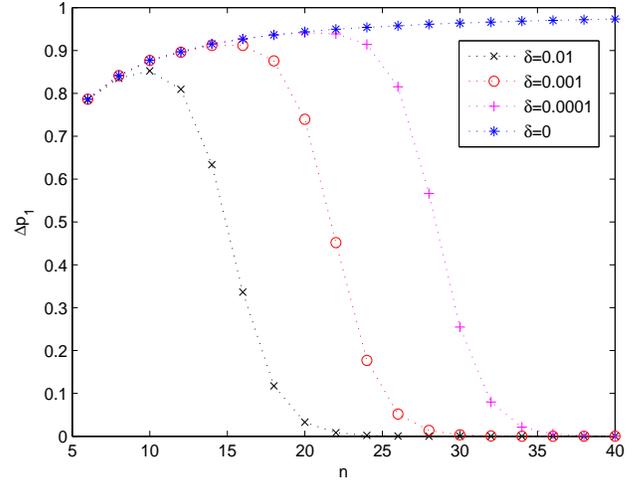}}

\caption{the probability gap of optimized SKW algorithm $\Delta{{p}_{1}}$ vs the scale of the database $n$ with different phase errors. The asterisks line for $\delta  = 0$, the ‘+’ line for $\delta  = 0.0001$ , the circles line for $\delta  = 0.001$ , the ‘x’ line for $\delta  = 0.01$ }
\label{fig:6}       
\end{figure}

\begin{figure}[htb]
 \resizebox{9.3cm}{7cm}{\includegraphics{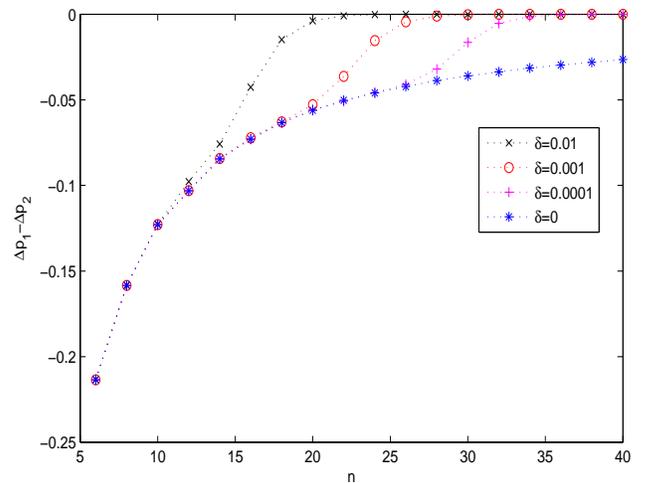}}

\caption{the difference between the probability gaps $\Delta{{p}_{1}}-\Delta{{p}_{2}}$ vs the scale of the database $n$ with different phase errors. The asterisks line for $\delta  = 0$, the ‘+’ line for $\delta  = 0.0001$ , the circles line for $\delta  = 0.001$ , the ‘x’ line for $\delta  = 0.01$ }
\label{fig:7}       
\end{figure}

There is a tendency that the probability gap of optimized SKW algorithm decreases exponentially with the increment of errors, shown in Fig. 6. But for the same size of databases the difference between the probability gaps increases gradually, as shown in Fig. 7. This means that the probability gap of Grover's algorithm decreases faster, and optimized SKW algorithm is more robust.

\section{Conclusion}
\label{sec:7}
This paper studies the effects of systematic errors in phase inversions on optimized SKW algorithm in detail. The geometric description of the algorithm is given out first. Then the model of this algorithm with phase errors can be built. It is found that phase errors will reduce both the number of iterations and the success rate of the algorithm through theoretical analysis and numerical simulations. The larger error will lead to the smaller maximum success rate of the algorithm and the fewer required iteration times. We obtain two expressions about the maximum success rate of the algorithm and the required iteration times, which are both related to the phase error and the database size. Then we obtain a critical value of the database scale ${n}$. When ${\rm{n}} \le 1.806 \cdot \log _2^{\frac{{0.4642}}{\delta }}$, the algorithm is almost unaffected by the error. Based on the above relationships, we depict the relationship between the success rate of the algorithm, the phase error, the number of iterations and the database size. Finally, Analysis and numerical simulations show that optimized SKW algorithm is more robust than Grover's algorithm.

\section*{ACKNOWLEDGMENTS}
 The authors gratefully acknowledge the financial support from the National Basic Research Program of China (Grant No. 2013CB338002).

\end{document}